\documentclass[sigconf,letterpaper]{acmart}
\usepackage{makecell}
\usepackage{placeins}
\usepackage{multirow}
\usepackage{float}
\usepackage{graphicx}
\usepackage{graphics}
\usepackage{subcaption}
\usepackage[skins]{tcolorbox}
\usepackage[ruled,vlined]{algorithm2e}
\usepackage{amsmath}
\usepackage{appendix}
\usepackage{url}
\usepackage{color}
\usepackage{breqn}
\usepackage{pifont}
\usepackage{algpseudocode,caption}
\usepackage{float}
\usepackage{threeparttable}
\usepackage[inline]{enumitem}
\usepackage{booktabs}
\newcommand{\etal}{{\em et al.}\xspace}
\usepackage{xcolor}
\newcommand{\BfPara}[1]{{\vspace{0.3em}\noindent\bf#1.}\xspace}
\usepackage{xcolor}
\usepackage{colortbl}
\usepackage{tabularx}
\colorlet{lightgrey}{lightgray}
\usepackage{tikz}
\usepackage{pgf}
\usepackage{float}

\AtBeginDocument{%
  \providecommand\BibTeX{{%
    \normalfont B\kern-0.5em{\scshape i\kern-0.25em b}\kern-0.8em\TeX}}}

\usepackage{amsmath}
\usepackage{graphicx}
\usepackage{makecell}
\usepackage{hyperref}
\newcolumntype{L}[1]{>{\raggedright\let\newline\\\arraybackslash\hspace{0pt}}m{#1}}
\newcolumntype{C}[1]{>{\centering\let\newline\\\arraybackslash\hspace{0pt}}m{#1}}
\newcolumntype{R}[1]{>{\raggedleft\let\newline\\\arraybackslash\hspace{0pt}}m{#1}}

\usepackage{tikz}

\setcopyright{none}
\renewcommand\footnotetextcopyrightpermission[1]{} 
\begin{document}

\title[Miners in the Cloud: Measuring and Analyzing Cryptocurrency Mining in Public Clouds]{Miners in the Cloud: Measuring and \\Analyzing Cryptocurrency Mining in Public Clouds}

\begin{abstract}
Cryptocurrencies, arguably the most prominent application of blockchains, have been on the rise with a wide mainstream acceptance. A central concept in cryptocurrencies is ``mining pools'', groups of cooperating cryptocurrency miners who agree to share block rewards in proportion to their contributed mining power. Despite many promised benefits of cryptocurrencies, they are equally utilized for malicious activities; e.g., ransomware payments, stealthy command, control, etc. Thus, understanding the interplay between cryptocurrencies, particularly the mining pools, and other essential infrastructure for profiling and modeling is important. 

In this paper, we study the interplay between mining pools and public clouds by analyzing their communication association through passive domain name system (pDNS) traces. We observe that 24 cloud providers have some association with mining pools as observed from the pDNS query traces, where popular public cloud providers, namely Amazon and Google, have almost 48\% of such an association. Moreover, we found that the cloud provider presence and cloud provider-to-mining pool association both exhibit a heavy-tailed distribution, emphasizing an intrinsic preferential attachment model with both mining pools and cloud providers. We measure the security risk and exposure of the cloud providers, as that might aid in understanding the intent of the mining: among the top two cloud providers, we found almost 35\% and 30\% of their associated endpoints are positively detected to be associated with malicious activities, per the \url{virustotal.com} scan. Finally, we found that the mining pools presented in our dataset are predominantly used for mining Metaverse currencies, highlighting a shift in cryptocurrency use, and demonstrating the prevalence of mining using public clouds.
\end{abstract}



\author{Ayodeji Adeniran}
\affiliation{%
   \institution{University of Central Florida}
   \city{Orlando}
   \state{FL}
   \country{USA}}
\email{aadeniran@Knights.ucf.edu}

\author{David Mohaisen}
\affiliation{%
   \institution{University of Central Florida}
   \city{Orlando}
   \state{FL}
   \country{USA}}
\email{mohaisen@ucf.edu}

\maketitle

\section{Introduction}\label{sec:introduction}
Cryptocurrencies have recently been on the rise, with the top three cryptocurrencies amounting to over a trillion USD in value~\cite{SaadCM21,SaadCNKM20}. With cryptocurrency gradually gaining acceptance, different cryptocurrencies have emerged and are still emerging~\cite{dziembowski2015introduction}. Many cryptocurrencies are a direct application of blockchain technology, a distributed ledger over a distributed network of nodes that record transactions~\cite{SaadCM21}. The blockchain distributed system provides a much safer architecture against failure, and that---along with cryptographic primitives---ensure transactions are safer from being altered~\cite{AhmadSMB18,AhmadSM19,meiklejohn2013fistful}. Given the importance and value of those cryptocurrencies, cybercriminals have used them as a way to enable their criminal activities~\cite{SaadSurvey20}.

Cybercrimes have been evolving over the years, where cybercriminals have been continuously coming up with new ways to violate system security properties, steal information, hijack resources, and demand ransom~\cite{kim2020detecting,baek2018ssd,mohaisen2013unveiling,RuthWH18,pastrana2019first,meiklejohn2013fistful,scaife2016cryptolock,chernikova2022cyber}. The emergence of new attacks has been a continuous race between the attackers and the defenders. The attackers have used several platforms to launch attacks. Attackers were noticed to have changed their strategy to defeat defenses~\cite{HuangDMDGMSWSL14}. For instance, with the emergency of blockchain-based technologies, malicious transactions are placed on the blockchain to facilitate malicious activities through the distribution of stealthy command and control channels~\cite{bock2019assessing}. Given the significant valuation of cryptocurrencies, cryptojacking, an intentional effort to use others' machines and resources for mining cryptocurrencies has been on the rise~\cite{SaadKM19}. Particularly, the use of cloud resources has been hypothesized to be the main entry point of mining cryptocurrencies~\cite{KharrazMMLMMBAB19}, although not systematically analyzed.

The focus of this work is to understand the prevalence of public clouds for mining purposes, possibly mining with malicious intent (e.g., with compromised cloud instance). We hypothesize that utilizing cloud resources for such activities is more consistent with the general compute trends, and (from a security standpoint) adversaries’ incentives than ever before. For instance, launching cyber-attacks from the private server(s) that can be traced and shutdown is no longer popular among attackers because, apart from the ease and flexibility of springing up resources in the cloud, they also make less return on investment compared to setting up private servers. There is a number of cloud providers, from enterprise-scale to small-scale, the users and adversaries alike can move from one provider to another and set up attack fronts fast.  Exploiting blockchain technology in conjunction with cloud resources, the attackers have a huge and cheap cloud to benefit from, and a difficult-to-decipher blockchain technology to hide their malicious activities, particularly when those cloud resources are obtained free of charge (i.e., compromised). While security analysis is a byproduct of our analysis, it also highlights the general trend in this space. 

In this paper, we study the interplay between mining pools and public clouds by analyzing their communication association through passive Domain Name System (pDNS) traces. We observe that 24 cloud providers have some association with mining pools as observed from the pDNS query traces, where two popular public cloud providers--Amazon and Google--have almost 48\% of such an association. We found that the cloud provider presence and cloud provider-to-mining pool association both exhibit a heavy-tailed distribution, emphasizing an intrinsic preferential attachment model with both mining pools and cloud providers. We measure the security risk and exposure of the cloud providers, as that might aid in understanding the intent of the mining: among the top two cloud providers, we found almost 35\% and 30\% of their associated endpoints are positively detected to be associated with malicious activities, per the \url{virustotal.com} scan. Finally, we found that the mining pools presented in our dataset are predominantly used for mining Metaverse currencies, highlighting a shift in cryptocurrency use, and demonstrating the prevalence of mining using public clouds.

\BfPara{Organization} The related work is presented in \autoref{sec:related}, followed by our dataset overview in~\autoref{sec:dataset}, the main results in~\autoref{sec:results}, the discussion in \autoref{sec:discussion}, and concluding remarks in \autoref{sec:conclusion}. 

\section{Related Work}\label{sec:related}

This work is broadly associated with a body of work on cryptomining (irrespective of the tools used for that mining). Tahir~\etal~\cite{TahirHDAGZCB17} studied the abuse of virtual machines in cloud services for mining digital currencies, which is the most related work to ours. Huang \etal \cite{HuangDMDGMSWSL14}  were among the first to notice the illegal use of CPU cycles for malware-induced mining. The initial work on web-based cryptomining was presented by Saad \etal~\cite{SaadKM19} and Ruth \etal~\cite{RuthWH18} who measured the prevalence of cryptojacking among websites (i.e., utilizing mining on visitors' machines). 

Ruth \etal~\cite{RuthWH18} obtained blacklisted URLs using the no coin web extension, mapped them on a large corpus of websites obtained from the Alexa Top 1M list, and identified 1491 suspect websites involved in cryptojacking. In a concurrent work, Saad \etal~\cite{SaadKM19} conducted a similar study, but on a larger number of websites; 5703 sites in total. Concurrently, Eskandari \etal~\cite{EskandariLMC18} examined the prevalence of cryptojacking among websites and the use of {\em Coinhive} as the most popular platform for cryptojacking. All of these studies highlight the issue of cryptojacking through measurements, and the emerging use of cryptojacking as an alternative to online ads. Saad \etal~\cite{SaadKM19} goes further by conducting code analysis toward the detection of cryptojacking codes and their economic impact. 

Bertino and Nayeem~\cite{BertinoN-17} highlighted worms in IoT devices that hijacked them for mining purposes, pointing to the infamous {\em Linux.Darlloz} worm that hijacked devices running Linux on Intelx86 chip architecture for mining~\cite{Bansal-14}. Krishnan~\etal~\cite{KrishnanSV-17} studied a series of computer malware, such as {\em TrojanRansom.Win32.Linkup} and {\em HKTL\_BITCOINMINE}, that turned host machines into mining pools. Sari and Kilik~\cite{SariS-17}, used Open Source Intelligence (OSINT) to study vulnerabilities in mining pools with Mirai botnet as a case study.

\section{Dataset and Preprocessing}\label{sec:dataset}

The dataset used in this study couples an enumeration of mining pools, and associations between them and cloud instances that belong to various cloud providers utilizing DNS query data. To this end, the first part of the input data used for this analysis is the mining pools and their associated addresses. The list of the mining pools was sourced by manually listing the mining domains from the Stelareum mining pool website. This list contains a set of mining pool addresses that are publicly available~\footnote{\url{https://www.stelareum.io/en/mining/pool.html}}. We examined the mining pools and copied the corresponding URLs for each of them. Subsequently, using the popular Digital Envoy IP allocation dataset\footnote{\url{https://www.digitalenvoy.com/}}, we enumerated the IP pools of some top public cloud providers, which formed the second input data. 

To establish an association between the various cloud providers and the mining pools, we conducted a scan over all the IPs allocated to the public providers on one side and the mining pools addresses on the other side using the passive domain name system (pDNS) dataset used in~\cite{PerdisciPAA20}. The scan utilizes the pDNS dataset to map a relationship between the mining pools and endpoints in the cloud where traffic is sent from those cloud endpoints to the mining pools, and vice versa, at some point in time in the past. The output of the scan consists of the originating IP address of the mining pool and the corresponding IP subnet of the cloud providers.

The scan result includes mining pool domain, pool source IP address, and public IP subnets. The IP subnets are then converted to their respective domains to get the name of the cloud providers. The data contains thousands of lines of response (as there could be various pDNS entries associated with the same pair of endpoints). As such, the data was cleaned by rearranging the data in descending order and then removing those with the least number of responses. To obtain the geographical distribution of the cloud providers in our data, we further augment the data with the country where the cloud provider is located.

\section{Results}\label{sec:results}
In this section, we present our main results by measuring and mapping the landscape of association between cryptocurrencies and public clouds. Before we dive into our analysis, we review the main dimensions of our analysis. 

\subsection{Analysis Dimensions}
In this study, we are concerned with various dimensions that highlight the interplay between public clouds and cryptocurrencies. Namely, we are concerned with cloud providers (associated with cryptocurrencies) and their geographical affinities, pool size, mining pools, and their distribution, cloud providers-specific distribution, and (potential) illicit activities. We define each of those dimensions as we present the associated results. 

\subsection{Results and Discussion}

\subsubsection{Cloud Providers and Country of Origin} There is a large number of public cloud providers that make up the cloud ecosystem. While the major providers are only a few (e.g., Amazon, Google, Azure, and Cloudflare), there are more than thousands of such providers. Understanding the affinity between mining pools and those cloud providers through distribution analysis is important for two reasons. First, such an analysis will allow us to understand the regional distribution of those providers and associated mining activities. Second, this analysis will further shed light on whether the large providers, in general, are still dominant in their use for mining. Answering this question, possibly positively, would allow us to devise effective policies to counter cryptomining threats. Moreover, insight for this analysis would draw a representative picture of the overall computing ecosystem and associated security characteristics.

\BfPara{Observations} \autoref{tab:main} shows the cloud providers mapped to countries of their domain registrations. Interestingly, we find that the distribution of the traffic from the mining pools is vastly distributed, covering a large number of providers and countries, and spanning several continents. Moreover, we found that the distribution of the cloud providers' representation with respect to the studied mining pools and their association is quite skewed (heavy-tailed) towards a few among them: while there are 24 different cloud providers represented in the dataset, the top 2 (Amazon and Google) have a representation of 48\%, while the next 12 providers have 42\%.

\autoref{tab:main} highlights the geographical distribution of the different cloud providers, where the distribution is also heavy-tailed over 15 countries, led by the US (57\%), followed by Russia (11\%), South Korea (6\%), and Japan (5\%). The remaining 11 countries have 21\% of the cloud endpoints shared among them collectively. The cloud distribution is obtained from the result of the pDNS scan. 

\begin{table}[t]
\centering
\caption{Cloud distribution across the country with the frequency of appearance.}\label{tab:main}
\begin{tabular}{L{0.17\textwidth}L{0.10\textwidth}R{0.06\textwidth}R{0.06\textwidth}}
\Xhline{2\arrayrulewidth}
Cloud Provider   & Country              & \# & \% \\
\Xhline{2\arrayrulewidth}
Amazon           & USA                  & 738        & 24\%    \\
Google           & USA                  & 737        & 24\%    \\
KORNET           & South Korea          & 188        & 6\%     \\
Cloudflare       & USA                  & 175        & 6\%     \\
Asia Pacific Net & Japan                & 161        & 5\%     \\
CL-KARELIA       & Russia               & 155        & 5\%     \\
SCL66-rented1    & Cyprus               & 135        & 4\%     \\
MACROREGIONAL    & Russia               & 117        & 4\%     \\
HIPL-SG          & USA                  & 69         & 2\%     \\
Corpori          & Brazil               & 58         & 2\%     \\
HOSTERION-SRL    & Romania              & 50         & 2\%     \\
KAZAKTELECOM     & Kazakhstan           & 46         & 2\%     \\
MOTIV-DC-1       & Netherlands          & 46         & 2\%     \\
MOTIV-DC-3       & Netherlands          & 46         & 2\%     \\
DNAP-081217      & Finland              & 44         & 1\%     \\
AOSOZVEZDIE-NET  & Russia               & 38         & 1\%     \\
IPNET-DS-WBS     & South Africa         & 33         & 1\%     \\
RS-KOPERNIKUS    & Rep of Serbia        & 33         & 1\%     \\
PS-1\_2177       & Kazakhstan           & 32         & 1\%     \\
BTC-TEMP1        & Bulgaria             & 31         & 1\%     \\
TR-RTNET-981210  & Turkey               & 31         & 1\%     \\
CLOUDFLARENET-EU & USA                  & 28         & 1\%     \\
RU-MOS-SMILE     & Russia               & 28         & 1\%     \\
UK-NTLI-990527   & UK       & 24         & 1\%     \\
\Xhline{2\arrayrulewidth}
                 & Total                & 3043       & 100\%   \\
\Xhline{2\arrayrulewidth} 
\end{tabular}
\end{table}

\subsubsection{Mining Pools and Associated Size} In this study, we measured two major pools in terms of their presence in the cloud. The size of the pool is measured by counting the number of the individual (cloud) IP addresses associated with it (i.e., issuing queries). Understanding the pool size would highlight which pool is more popular and central in the cryptocurrency ecosystem, and possibly which cryptocurrency is being mined by the pool utilizing cloud resources. 

\begin{table}[t]
\centering
\caption{Mining pools distribution. A heavy-tailed distribution in terms of the number of public cloud associated with the different mining pool (sub)domains.}\label{tab:mining}
\begin{tabular}{L{0.22\textwidth}R{0.1\textwidth}R{0.1\textwidth}}
\Xhline{2\arrayrulewidth}
Pools (subdomain)      & \# & \% \\
\Xhline{1\arrayrulewidth}
\url{sandpool.org}           & 1,470        & 41\%          \\
\url{etp.sandpool.org}       & 1,205        & 34\%          \\
\url{eu.miningethereum.net}  & 351         & 10\%          \\
\url{miningethereum.net}     & 348         & 10\%          \\
\url{www.miningethereum.net} & 73          & 2\%           \\
\url{www.sandpool.org}       & 55          & 2\%           \\
\url{ru.etp.sandpool.org}    & 26          & 1\%           \\
\url{dev.sandpool.org}       & 20          & 1\%           \\
\Xhline{1\arrayrulewidth}
Total & 3,043 & 100\% \\
\Xhline{2\arrayrulewidth}
\end{tabular}
\end{table}

\BfPara{Observations} We emphasize that we conducted scans of several mining pools in our initial data gathering (i.e., all those present in our initial set), although we only got a response from two mining pools domains, \url{sandpool.org} and \url{miningetherium.net}---which means that the other mining pools did not have any association with public clouds. In our scan, we noticed that the two mining pool domains contain other subdomains which responded to the query from the pDNS. \autoref{tab:mining} shows the mining domain and the subdomain with the corresponding representation in terms of cloud presence. Based on these results, we narrow down the focus of this paper to the two pools and their associations with the public cloud. 

Cryptocurrency mining has global acceptance with mining activities being carried out in several parts of the world. Some mining pools are rated to be in the top tier because of the amount of mining traffic recorded and associated with them. These pools are located in different countries. Most of the top-rated pools are located in China. For instance, in the general cryptocurrency ecosystem, pools like F2Pool, AntPool, BTCC, and BW account for more than 60\% of all the new bitcoins. While the dataset from the DNS scan recorded most traffic to Amazon and Google cloud providers, a search on the reported top 10 mining pools using censys.io, a search engine that scans the internet for connected devices. Alibaba's cloud network recorded a higher rate of traffic for the mining pools based in China, while Amazon had higher traffic for pools located in the US and in other countries. The geographical location of the mining pool could be a factor in determining the preferred cloud provider before selecting other available providers. 

By the same token, we also found that the top two subdomains (corresponding to the sandpool mining pool) represent 75\% of the overall cloud associations, with ETP, the second-largest association, representing 34\% of the associations (1,205 cloud endpoints). Upon further exploration, we found that this pool is used mostly for mining Metaverse ETP, a cryptocurrency that powers the Metaverse blockchain-as-a-service (BAAS) platform and is located in Europe. Among those two pools, sandpool.org had 78\% (or 2,750) of the cloud associations overall, while miningetherium.net had 22\% (or 798) of the cloud associations in total.

\begin{table}[htb]
\centering
\caption{Google (G) vs Amazon (A)  Cloud distribution between the two pools. While the general trend of heavy-tailed distribution still applies to the individual cloud shares against the subdomains of the pools, Amazon has a more skewed distribution in contrast to a more evenly distributed share of Google's cloud instances. }\label{tab:gvz}
\begin{tabular}{L{0.20\textwidth}R{0.04\textwidth}R{0.04\textwidth}R{0.04\textwidth}R{0.04\textwidth}}
\Xhline{2\arrayrulewidth}
Pools                  & G & A & G\% & A\% \\
\Xhline{2\arrayrulewidth}
\url{etp.sandpool.org}       & 307    & 213    & 42\%     & 36\%     \\
\url{sandpool.org}           & 186    & 357    & 25\%     & 59\%     \\
\url{miningethereum.net}     & 106    & 9      & 14\%     & 1\%      \\
\url{eu.miningethereum.net}  & 104    & 5      & 14\%     & 1\%      \\
\url{www.miningethereum.net} & 19     & 3      & 3\%      & 0\%      \\
\url{www.sandpool.org}       & 7      & 9      & 1\%      & 1\%      \\
\url{ru.etp.sandpool.org}    & 5      & 6      & 1\%      & 1\%      \\
\url{dev.sandpool.org}       & 3      & 5      & 0\%      & 1\%      \\
\Xhline{2\arrayrulewidth}
Total                  & 737    & 607    &    100     &100   \\
\Xhline{2\arrayrulewidth}
\end{tabular}
\end{table}

\subsubsection{Cloud Providers Distribution vs Pools} The current cloud ranking by market share places Amazon with the largest share, followed by Microsoft Azure, then Google. In our measurement, we found that Amazon and Google represented almost fifty percent of the total cloud instances connected to the mining pools in our dataset, while the other different cloud providers make up the remaining half. We want to further understand the detailed distribution between the two major cloud providers to the two mining pools to characterize their similarities and differences, as depicted in \autoref{tab:gvz}.

\BfPara{Observations} From those results, we made two key observations. First, the per-cloud distribution follows a similar trend of heavy-tail as in the generation distribution, although less skewed in the case of Google, where cloud share is distributed more evenly on a larger number of pool domains. Second, while Azure is quite popular in the abstract, and on par with the popularity of Google Cloud, it is absent from this analysis. We still speculate that cloud popularity may have played a factor in the distribution, and the absence of Microsoft Azure cloud in the distribution would possibly point to other factors that may have influenced the selection of Amazon and Google (e.g., strict abuse policies, or the popularity of this cloud in a given country). Some smaller cloud providers also reported traffic from the pool that indicated association to the clouds that could be either randomly or selectively. The information in the dataset is not explicit enough to accurately provide the details, but we hypothesize that factors like cost, security, or restrictions for some category of users, such as in the case of Azure, could be responsible for the choice and the clear trend.

\subsubsection{Malicious Associations} The popularity of cloud services over the years has made them attractive for both benign and malicious use. Services and applications previously hosted on private servers are now hosted in the cloud and many public and private companies are still migrating their workloads to the cloud. It is no surprise that mining activities are shifting to the cloud, considering the cost and the flexibility offered. Setting up servers in the cloud takes a few minutes at a significantly lower cost. The flexibility of moving from one cloud provider to another could be another factor besides the cost that attracted the miners to shift their activities to using cloud resources.

Cybercriminals operate by masquerading their malicious traffic from detection using different techniques. The servers are shut down once detected, which is a big loss for cybercriminals. Cloud services provide an easy solution for cybercriminals addressing this issue.  For instance, to make their activities more discreet, cybercriminals make use of blockchain technology when operating from the cloud. In case of detection, they quickly move to another cloud provider to set up their servers in a few minutes with very minimal disruption to their services and activities. They continue to operate in these cycles at a relatively small cost and manage to keep afloat for a while before being detected. 

A central question in our analysis is whether some of those cloud instances used for mining cryptocurrencies are involved in malicious activities. Unfortunately, the dataset we have is limited in many ways, particularly the absence of a payload for the DNS resolution or subsequent application-layer communication, which makes it impossible to draw such a conclusion. However, utilize our knowledge of the endpoint on the cloud to frame the question into a plausibility analysis: among those IP addresses associated with the cloud providers, how many of them are associated with malicious activities?

\BfPara{Observations} In order to address this question, we conducted an additional scan on the IP addresses associated with the cloud instances using \url{virustotal.com}, which is one of the three online scanning sites we used in our analysis. The results of the scan are shown in Table~\ref{tab:negative}. Among the 24 cloud providers reported in \autoref{tab:main}, only five cloud providers have positive scan results in \url{virustotal.com}, namely Amazon, Google, KORNET, Cloudflare, and CL-KARELIA. Among them, CloudFlare had the largest detection rate, with 44\% of the cloud instances reported by VirusTotal as having some security issues (i.e., flagged as a source of malicious activity). The percentage is followed by Amazon (at 34.69\%) and Google (29.85\%). Among those cloud providers KORNET had the least positive rate, with only around 1\% of the instances detected by \url{virustotal.com}.

While those results are inconclusive, and cannot be used to argue for intent associated with the mining activities taking place on those cloud providers for the different mining pools, or whether mining is taking place if at all, the fact that a positive detection is associated with a number of those public cloud IPs highlights the potential risk associated with those instances.

\begin{table}[]
\centering
\caption{A distribution analysis of the malicious cloud instances (IPs) in contrast to the total number of IP count associated with the given cloud provider, and the associated percentage. Notice that with the top cloud providers, a significant number of instances are shown to be associated with a malicious activity at some point in time, per \url{virustotal.com} scan. Cloud providers not present in this table returned negative scan results.}\label{tab:negative}
\begin{tabular}{{L{0.11\textwidth}R{0.08\textwidth}R{0.08\textwidth}R{0.08\textwidth}}}
\Xhline{2\arrayrulewidth}
Cloud Provider & Count & Malicious & \% \\
\Xhline{2\arrayrulewidth}
Amazon         & 738   & 256       & 34.69  \%    \\
Google         & 737   & 220       & 29.85 \%      \\
KORNET         & 188   & 2         & 1.06 \%      \\
CloudFlare     & 175   & 77        & 44 \%        \\
CL-KARELIA     & 161   & 10        & 6.21 \%     \\
\Xhline{2\arrayrulewidth}
\end{tabular}
\end{table}

\section{Discussion}\label{sec:discussion}

From our analysis, we notice that more associations are reported on the Google cloud platform than on the Amazon platform. The preference of Google Cloud over Amazon by the miners could be due to various reasons. For instance, computing power and the cost of electricity are among the challenges in cryptomining. Both cloud providers have instances with computer power resources to handle the cryptomining activities, but the cost might be the main differentiating factor because the profit determines the attractiveness of mining. The two top cloud providers are known to have high reliability and availability, and provide a range of offerings that suit the applications associated with the highlighted mining activities; virtual reality. Given the expectations of those applications, the failure rate is low in these clouds and miners could run their mining system uninterrupted.

By the same token, we observe that the less popular and smaller the cloud providers have fewer resource offerings and may lack the capacity required by cryptominers. Cybercriminals using blockchain for sending malicious traffic might prefer smaller cloud providers because they may be cheaper, prone to exploits, and less secure in general, while they may accommodate such activity to drive traffic on their cloud platform. There is the possibility of using popular clouds as well by exploiting the vulnerabilities in such clouds, especially by hijacking user accounts with weak credentials and hiding their malicious traffic among millions of packets originating from the cloud. Our security analysis above highlights the potential of this hypothesis, as a number of public cloud nodes (identified by their addresses) are shown to be associated with malicious activities in \url{virustotal.com} scan. 

Security in the cloud is a shared model system whereby both the cloud provider and the customer have their responsibilities shared. While several security measures are recommended, some fail to implement the required minimum security and may have their account hijacked and used for malicious purposes, including mining cryptocurrency. Hijacking user accounts for mining activity is common because of the high compute resources required for mining and this comes with a price: using someone else's account in this manner transfers the cost to the account owner while the miners earn rewards for their mining activity. These kinds of account hijacking for mining activities were reported by a number of cloud providers, especially by the top providers. For instance, 86\% of the hijacked accounts in Google clouds are used for cryptomining~\cite{cloudjacking}. Our results allude to a similar outcome, as many of the cloud addresses shown in our analysis are associated with malicious activities. 

In this study, we noticed that the two main mining pools uncovered in our analysis are quite protective of registration information. Upon digging into the DNS records of their domains, we found that their DNS resolution and hosting are done by Cloudflare. Neither of those main pools, nor their associated subdomains, are detected by \url{virustotal.com}. To contrast this result with other major mining pools, we evaluated the top mining pools (besides those studied thus far). The results are shown in Table~\ref{tab:top10}. We noticed all the top ten mining pools are hosted in the cloud as well, and mostly in Cloudflare. Out of the ten mining pools, nine are hosted on the Cloudflare cloud, highlighting a persistent trend in the utilization of cloud resources for running mining pools, perhaps for their high availability. The public cloud provides some security measures, but that does not necessarily prevent the pools from being used for malicious purposes.  We then scanned all the mining pools IP addresses using \url{virustotal.com}. While all of them returned negative detection results (although many returned ``unrated'' result for the scanned addresses), indicating that they were not involved in reported malicious activities. However, two types of detection were reported: passive DNS (pDNS) replication and communication file detection.

Interestingly, none of the domain names associated with those pools have pDNS flag; pDNS is what we use for retrieving the association between clouds and mining pools. The pDNS Replication provides temporary storage for DNS queries and captures the queries on the network and stores them for later retrieval. The stored queries are in historical form, which can be analyzed later by security experts. \url{virustotal.com} explains the main idea behind passive DNS as inter-server DNS which captures messages and forwards them to a collection point for analysis and storing of the individual DNS records in a database where they are indexed and queried after the processing. Given the lack of pDNS data for those domains, it is not surprising that we could not see them in our initial association dataset. To this end, our analysis comes with a caveat: the estimated association between those cloud providers and mining pools is a lower bound, and only captures those that are explicit about their association. 

The files entry in~\autoref{tab:top10} highlights the number of files that have been determined to perform some kind of communication with the IP address of the domain under consideration. These files are not considered malicious in nature, but indicative of an association with other addresses, which highlights our hypothesis that the estimated association is a lower bound.

\begin{table}[t]
\centering
\caption{Top 10 mining pools scan. IP addresses are masked for privacy.}\label{tab:top10}
\begin{tabular}{L{0.13\textwidth}L{0.08\textwidth}L{0.08\textwidth}R{0.05\textwidth}R{0.05\textwidth}}
\Xhline{2\arrayrulewidth}
Mining Pool & IP Address & Domain & pDNS & Files \\
\Xhline{2\arrayrulewidth}
\url{Binance.com} & 13.226.**.** & Amazon & 0 & 2 \\
\url{slushpool.com} & 104.26.**.** & Cloudflare & 0 & 1 \\
\url{f2pool.com} & 104.18.**.** & Cloudflare & 0 & 10+ \\
\url{pool.btc.com} & 104.18.**.** & Cloudflare & 1 & 10+ \\
\url{viabtc.com} & 104.16.**.** & Cloudflare & 0 & 3 \\
\url{v3.antpool.com} & 104.18.**.** & Cloudflare & 0 & 4 \\
\url{poolin.com} & 104.22.**.** & Cloudflare & 0 & 5 \\
\url{bw.com} & 172.66.**.** & Cloudflare & 0 & 7 \\
\url{bitfury.com} & 104.26.**.** & Cloudflare & 0 & 10+ \\
\url{v3.antpool.com} & 104.18.**.** & Cloudflare & 0 & 4\\
\Xhline{2\arrayrulewidth}
\end{tabular}
\end{table}

\begin{table}[htb!]
\centering
\caption{Mining Pool IP address Malicious scan.}\label{tab:malicious}
\begin{tabular}{L{0.15\textwidth}L{0.18\textwidth}R{0.07\textwidth}}
\Xhline{2\arrayrulewidth}
Pool & Security Vendor & Number \\
\Xhline{2\arrayrulewidth}
\url{v3.antpool.com} & CMC Threat Intelligence & 1 \\
\url{bitfury.com} & CMC Threat Intelligence & 1 \\
\url{poolin.com} & CMC Threat Intelligence & 1 \\
\url{v3.antpool.com} & CMC Threat Intelligence & 1 \\
\url{viabtc.com} & CMC Threat Intelligence & 1 \\
\url{pool.btc.com} & CMC Threat Intelligence & 1\\
\Xhline{2\arrayrulewidth}
\end{tabular}
\end{table}

The analysis we have conducted thus far is based on \url{virustotal.com}, which is the golden standard for evaluating security through detection against a range of scanners and antivirus vendors. We scan the pools and associated subdomains using various in-house products of threat intelligence. Interestingly, and as shown in Table~\ref{tab:malicious}, we found that a number of those pools are reported as involved in malicious activities (in the description,  CMC threat intelligence reported that malware is hosted on the listed mining pools). While a detection that is not replicated by the other major vendors in \url{virustotal.com}, highlights a divide in the security industry on how mining is perceived.

\section{Conclusion}\label{sec:conclusion}
In this paper, we initiate the systematic study between public clouds and cryptocurrencies, one of the most prominent applications of blockchain systems. Through pDNS traces, we establish the association between two mining pools and cloud providers. Unsurprisingly, we found that the major cloud providers are popular in their association with mining pools, with a heavy-tailed distribution and global presence. Upon examining the security of the associated cloud endpoints associated with mining pools, we found that a significant number of them (above 30\% in three cases) are malicious by \url{virustotal.com} scan results. By examining the hosting patterns of mining pools, we found that they are also heavily utilizing cloud providers, and the view of those mining pools, from a security standpoint, is divided. While our study is limited by the lack of payload from which one could understand the intent of the different associations between cloud and mining pools, it calls for further actions in this direction by providing preliminary anecdotes through characterization. 

\subsection*{Acknowledgement} This work is supported in part by NRF grant NRF-2016K1A1A2912757. We would like to thank Omar Alrawi (Georgia Tech) for providing us with the passive DNS dataset used in this study, and for the various insightful discussions.


\end{document}